# Drawing materials studied by THz spectroscopy

Andrea Taschin[1,2], Paolo Bartolini[1], Jordanka Tasseva[3], Jana Striova[4], Raffaella Fontana[4], Cristiano Riminesi[5] and Renato Torre[1,2,*].

[1] European lab. for Non-Linear Spectroscopy (LENS), Univ. di Firenze, via N. Carrara 1,I-50019 Sesto Fiorentino, Firenze, Italy.
[2] Dip. di Fisica e Astronomia, Univ. di Firenze, via Sansone 1, I-50019 Sesto Fiorentino, Firenze, Italy.
[3] INFN, Istituto Nazionale di Fisica Nucleare, Sez. di Napoli, Complesso Univ. di M. S. Angelo, Ed. 6- Via Cintia, 80126 Napoli.
[4] Istituto Nazionale di Ottica, INO-CNR, Largo Fermi 6, I-50125 Firenze, Italy.
[5] Istituto per la Conservazione e la Valorizzazione dei Beni Culturali, ICVBC-CNR, Via Madonna del Piano 10, I-50019 S Sesto Fiorentino, Italy.

*\* e-mail: torre@lens.unifi.it*

ABSTRACT

THz time-domain spectroscopy in transmission mode was applied to study dry and wet drawing inks. In specific, cochineal-, indigo- and iron-gall based inks have been investigated; some prepared following ancient recipes and others by using synthetic materials. The THz investigations have been realized on both pellet samples, made by dried inks blended with polyethylene powder, and layered inks, made by liquid deposition on polyethylene pellicles. We implemented an improved THz spectroscopic technique that enabled the measurement of the material optical parameters and thicknesses of the layered ink samples on absolute scale. This experimental investigation shows that the THz techniques have the potentiality to recognize drawing inks by their spectroscopic features.

## 1. INTRODUCTION

During the last years, THz spectroscopy has been intensely developed thanks to the realization of new laser sources capable of generating coherent radiation. Because the THz radiation is of low energy, THz spectroscopic techniques are non-destructive tools able to provide valuable information in the cultural heritage field. Recently, some studies [1-3] have demonstrated the possibility to use this radiation in the investigation of painting and drawing media. In particular, the THz-Time Domain Spectroscopy (THz-TDS) proved to be a powerful THz spectroscopic technique able to measure different optical parameters of the artistic works. These pulsed THz techniques can be implemented by using both the transmission or reflection configurations [4-7].

When a THz pulse passes through the sample, the material induces modifications in its temporal waveform; from the analysis of these variations the optical parameters of the material can be recovered. Nevertheless, the extraction of the transmission parameters (i.e. absorption coefficients and indexes of refraction) from the THz pulse waveform is not a trivial task [8]. It dependents on the experimental conditions and sample characteristics; in particular, the procedure becomes very complex for thin film samples [9].

Recently, we have presented a THz spectroscopic study of layered ink samples [7]. We implemented an innovative experimental and analytical method enabling a reliable extraction of the THz parameters from a TDS experiment. We report here on the investigation of a series of black and colored inks by means of this innovative THz-TDS approach. We measured both bulk inks in pellet samples and inks layered as thin film samples. Moreover, we compared the ink THz spectra with those of the separate specific constituents of the ink pigments. This enables one to identify the origin of the spectroscopic peaks present in the THz absorption spectra of the inks.

## 2. MATERIALS AND METHODS

### 2.1. Samples preparation

We investigated samples prepared in lab using home-made and commercial inks. The former inks have been prepared using both ancient recipes (Giovanni Alcherio 1411) and synthetic materials. Four groups of inks, according to the colour they render, might be distinguished: red, blue, black and white. In Table 1, we report the compositions of the studied inks and the recipes followed for their preparations. The black inks, all iron gall based, were prepared in lab following two main recipes containing either oak galls as a source of the gallo-tannic acid (Recipe A in Table 1) or preliminarily synthesized gallic acid (Recipe B in Table 1). Details about the ink preparation are reported in [7].

Dried inks were blended with polyethylene (PE) powder (Merck), ground, and pressed in pellets of 13.2 mm diameter and thickness around 1 mm. PE is the ideal support for absorption spectroscopy in the THz region thanks to its negligible absorption coefficient (below of 1 cm$^{-1}$ see [10] and references therein). The analyte concentration in the PE pellet was chosen to be approximately 33 wt. % that, for most of the studied inks, is an optimal concentration, rendering detectable any eventual features in a relatively wide spectral range. As an attempt of getting closer to a real-practice experiment on ancient manuscripts and drawings, some of the above inks have been studied also in the form of thin films, of the order of tens μm,

Table 1. Compositions and recipes of the studied inks.

| Color Ink | Ink type | Ink Recipes |
|---|---|---|
| Red | 1 | **Commercial cochineal carmine**: possibly based on carminic acid ($C_{22}H_{20}O_{13}$). |
|  | 2 | **Red ochre**: containing iron(III)oxide-hydroxide. |
| Blue |  | **Commercial blue ink**: based on 2,2-Bis(2,3-dihydro-3-oxoindolyli-den), Indigotin. |
| Black | A | **Recipe A - Iron gall ink with Arabic gum**: 70 mL water, 10 mL white wine, 10 mL red vinegar, 5 g powdered oak galls (Bizzarri, Firenze), 1.25 g Arabic gum, 1.25 g $FeSO_4 \cdot 7H2O$ (Bizzarri, Firenze) melanterite. |
|  | B | **Recipe B - Iron gall ink with Arabic gum**: 14 mL distilled water, 1.14 g gallic acid (Bizzarri, Firenze), 0.29 g Arabic gum, 0.29 g $FeSO_4 \cdot 7H2O$ (Bizzarri, Firenze). |
|  | C | **Commercial iron gall ink (Zecchi).** |
| White |  | **White lead**: $2PbCO_3 \cdot Pb(OH)_2$. |

layered on 10 µm PE pellicles (polyethylene far-IR sample cards by Sigma-Aldrich).

## 2.2. Terahertz time domain spectroscopy

Standard THz-TDS set-up in transmission configuration has been used to extract the values of the refractive index, $n$, and the absorption coefficient, $\alpha$, of inks in the frequency range 0.1 - 4 THz. THz pulses are produced by exciting a biased photoconductive antenna with femtosecond optical laser pulses ($\lambda$ = 780 nm, $\Delta t$ = 120 fs at 100 MHz). The outgoing THz radiation is guided, as usual, by two couples of off-axis parabolic mirrors. The THz field is collimated, focused on the sample, collimated again, and finally focused on a second photoconductive antenna for its detection. A second optical laser pulse, sent to the detection antenna, acts as the current gate for the THz detection. The temporal evolution of the photocurrent amplitude in the detection antenna, acquired by changing time delay between exciting and gate pulses, is directly related to the electric field amplitude of the THz radiation. This current is amplified by a lock-in amplifier and digitized by an acquisition board. The processed signal and delay line encoding are acquired by homemade software for retracing the final time dependent THz field. The whole THz set-up is enclosed in a nitrogen purged chamber for removing the water vapour contribution present at the THz frequencies spanned by the experiment.

More details about the used THz-TDS set-up can be found in reference [7] and the supplementary information therewith.

## 2.3. Materials optical parameters extraction

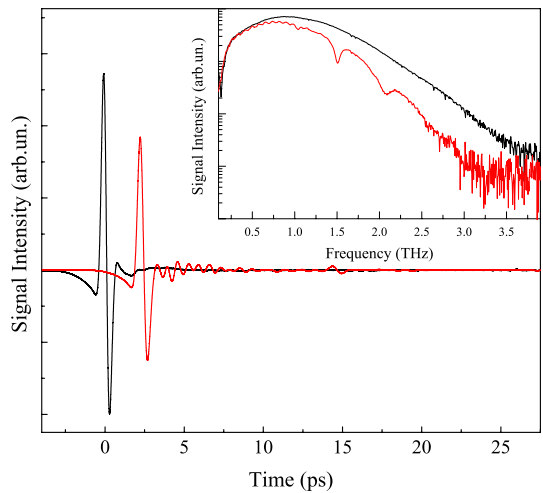

Figure 1. Main panel: typical time evolution of the THz electric field of the reference (black line) and signal (red line) in the Black B pellet sample. Inset: amplitude spectra of the signals obtained by their Fourier transforms.

The transfer function of the material, $H(\omega)$ (see eq. (1) in ref [7]) is the ratio between the THz field transmitted after the sample, $E_t(\omega)$, and the incident field, $E_i(\omega)$. It expresses how the phase and amplitude of an electromagnetic plane wave of frequency $\omega$ change due to the absorption and refraction of the crossed medium. The optical properties of the material, $n(\omega)$ and $\alpha(\omega)$, and its thickness, $d$, in principle, can be completely characterized by measuring the experimental transfer function, $H_{exp}(\omega)$, obtained by the ratio of the complex Fourier transform of the sample and the reference signal. Unfortunately, the expression for $H(\omega)$ is not expressed in a closed-form and the real thickness of the sample is not generally known with enough accuracy. Some numerical iterative process has to be used to extract $n(\omega)$, $\alpha(\omega)$ and $d$ [11-15] and this process depends on the nature of the THz signal and the thickness of samples.

Figure 1 reports typical signals measured by our THz-TDS set-up on Black B pellet sample. Both the temporal evolution of the THz signal (main panel of Figure 1) and its spectral content (inset of Figure 1) show modifications due to the absorption of the ink. The second pulse at around 15 picoseconds in time evolution of the sample THz signal is due to the internal reflections of the THz pulse between the sample surfaces. These appear in the spectrum as fast oscillations in the low frequency window.

If the reflections are clearly distinguishable in the temporal data and they are not superimposed on the oscillating modifications due to the sample absorption (contrary to what observed in Figure 1), the analysis for extraction of the optical parameters is relatively simple [7]. Usually, this occurs with thick samples and weak absorption.

Contrarily, in thin-layered sample, the reflection signals are close in time and partially superimposed, so a correct extraction of the material parameters requires a complex data analysis where the multi-reflection processes are properly taken into account [7-8]. The transfer function becomes more complex and the material parameters, $n(\omega)$ and $\alpha(\omega)$, can be extracted only using an articulated data analysis based on iterative fitting procedures [7-9]. Otherwise, the refractive index and absorption coefficient spectra are distorted by fake oscillations.

Recently, we have implemented an innovative experimental procedure and numerical method to analyse the transmission THz-TDS signal of samples composed of multiple thin layers [7]. This data analysis enables a reliable and robust extraction of the frequency dependence of the optical parameters, opening the possibility to measure even weak and broad peaks present in the $n(\omega)$ and $\alpha(\omega)$ functions. Moreover, it allows measuring the sample layer dimensions, even when it is no sufficiently thick to generate visible reflections in the time-domain data.

In this experimental investigation, we applied this method on the THz-TDS data from a double layered sample: an ink film deposited on polyethylene (PE) pellicles.

All details about the data analysis on single and dual thin layers are reported in ref. [7] and the supplementary information enclosed therewith.

## 3. RESULTS AND DISCUSSION

Due to the complexity in calculating the molar concentration of inks in pellets, only relative molar absorption coefficient and refractive index may be extracted and these are reported in relative units (Figures 2-6). For the pellet samples, scattering processes due to inhomogeneities in the sample could also affect the optical properties. In particular, the mismatch between the ink and PE indices gives rise to a spectral distortion called Christiansen Effect [16]. The effect can be greatly reduced by a fine grinding of the powders. Usually, the scattering introduces an additional absorption contribution, which grows with frequency, and distorts the absorption peaks and their dispersion curves. Yet, new fake absorption peaks could arise, as a results of interference of Mie scattering that, however, do not exhibit the characteristic dispersion in the refractive index. As long as the peaks in the measured absorbance spectra are of a symmetric Lorentzian line shape and are also present in the relative dispersions of the refractive index, the scattering processes can be considered negligible.

For the films, instead, the employed analysis allows the measurement of the optical properties of the same ink layer, and the absolute absorption coefficients and refractive indices are reported (Figure 7).

In Figure 2 we report the absorbance spectra of Red 1, Red 2, Blue, and White inks in pellets. Only Red 1 shows featureless spectra in the measured THz range whilst the others present a few spectral peaks. In particular, Red 2 shows a peak around 2.2 THz, Blue two very weak peaks at 1.5 and 1.95 THz and the White ink a broad band centred at 2 THz. These features are confirmed by the dispersion sample visible in Figure 3.

Figures 4 and 5 show the absorbance spectra and frequency dependence of the refractive indexes for all three black inks measured on pellets. The iron gall inks prepared following the old recipe of mixing mainly iron(II) sulfate and oak galls powder, Black A, shows a featureless response in the THz range. On the other hand, the ink containing synthetic gallic acid, Black B, and the commercial iron gall ink, Black C, show structured spectra with several peaks. Our results are in perfect agreement with what reported by Bardon et al. [6]. In order to ascribe each spectral component of Black B and Black C, we compared them with those of the individual constituents, gallic acid, iron(II) sulfate commercial, and iron(II) sulfate synthesized in lab (Figure 6). Gallic acid shows well-defined bands around 1.04, 1.50, 2.06, 2.57 and 3.09 THz. In the spectra of commercial ferrous sulfate, features at 1.12, 1.52, 1.92 THz are present. For the in-lab synthesized ferrous sulfate, instead, the peak at 1.12 THz disappears and an additional band around 3.01 THz is detected. The iron gall inks prepared following the recipe A do not show any features related either to gallic acid or to iron(II) sulfate. It is known that under hydrolysis gallic acid is extracted from the oak galls. Most likely all ferrous sulfate has oxidized to ferric sulfate, resulting in the formation of iron-tannic complexes. That hypothesis is well supported by Bardon et al. [6], that report no contribution of tannic acid in the 0.15 – 3 THz range and a single, questionable feature in the spectrum of iron(III) sulfate.

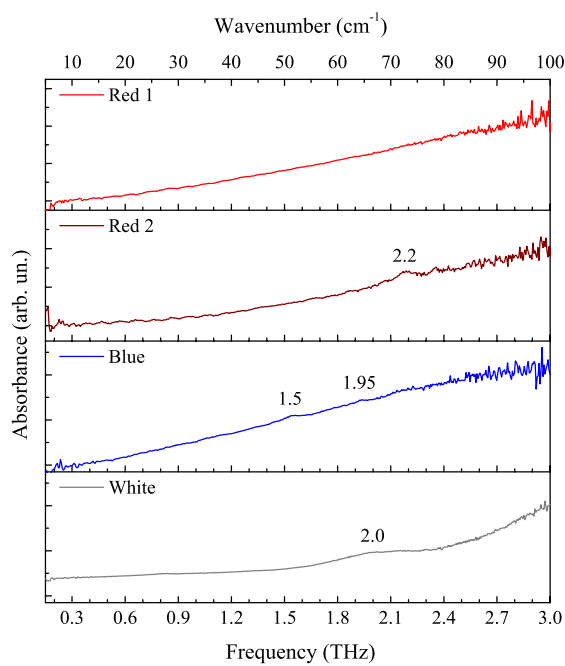

Figure 2. THz absorbance spectra for Red 1, Red 2, Blue, and White pigments measured on pellets. Some weak spectral contributions are evident in all the inks except in the cochineal carmine ink.

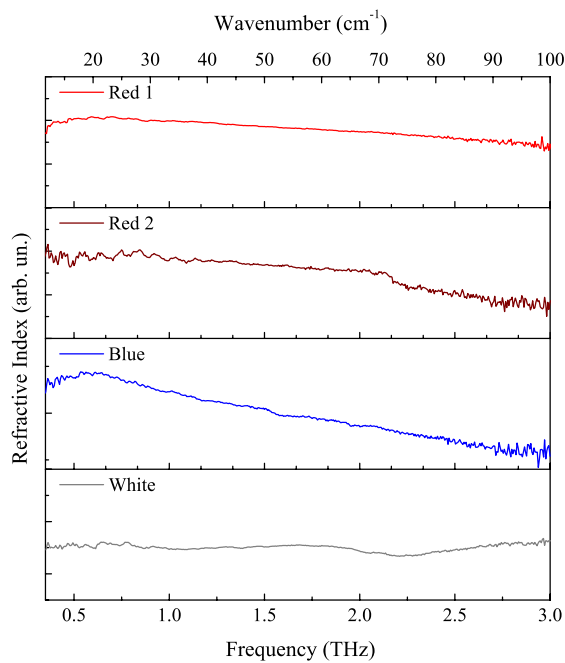

Figure 3. Frequency dependence of the refractive indices for Red 1, Red 2, Blue, and White pigments measured on pellets. The dispersion behaviour confirms the absorptive origin of the features found in the absorbance spectra of the Red 2, Blue, and White samples in the 1.5-2.5 THz range.

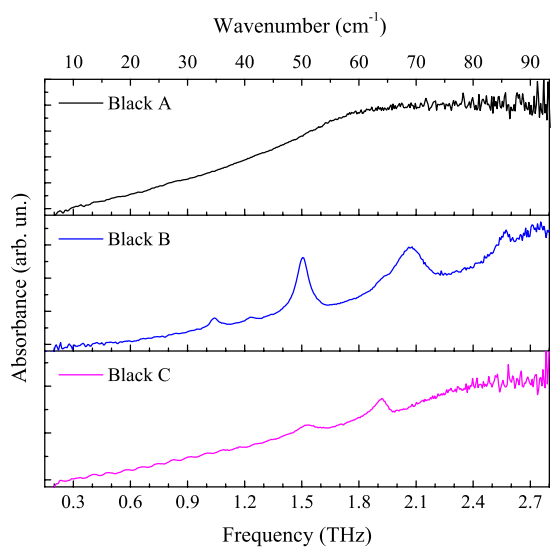

Figure 4. THz absorbance spectra for the iron gall inks measured on pellets. The pigments prepared following the ancient recipe (Recipe A) containing oak galls (Black A) do not show any particular spectral feature contrary to what observed in the inks containing synthesized gallic acid (Black B and Black C) which instead shows well pronounced peaks.

On the other hand, if gallic acid was extracted from the oak galls, and iron sulfate to gallic acid molar ratio was close to 1, the two components might have been consumed in the formation of iron gallate complexes, possibly featureless in the studied range. In the spectra of the lab-made iron gall ink prepared from ferrous sulfate and synthesized gallic acid (recipe B, Table 1), we can see the response of the gallic acid, while no ferrous sulfate contribution is present (Figure 6). Following what reported by other researchers [6], we could attribute this to the low iron sulfate to gallic acid molar ratio that in our case is 0.1. The commercially available iron gall ink, instead, shows features of the iron sulfate and lacks those of gallic acid as reported in Figure 6.

As a next step we decrease the analyte quantity, preparing thin films on transparent PE pellicles. The analysed inks were: the three black inks, Black A, Black B, and Black C and their optical parameters are shown in Figure 7. We were able to obtain also in this case a very good signal-to-noise ratio and to get agreement with the results from bulk samples. The bilayer analysis procedure [7] enables us to measure thickness down to 10 μm. Iron-gall ink, prepared with oak galls powder, once again does not manifest features of either gallic acid or ferrous sulfate. Similarly to the pellet sample, the layered iron-gall ink prepared following recipe B reveals the presence of gallic acid by peaks in absorbance at 1.04, 1.50, 2.06 and 2.57 THz. The film from commercial iron-gall ink shows the 1.92 THz feature of ferrous sulfate, while the one at 1.52 THz seems to be strongly damped.

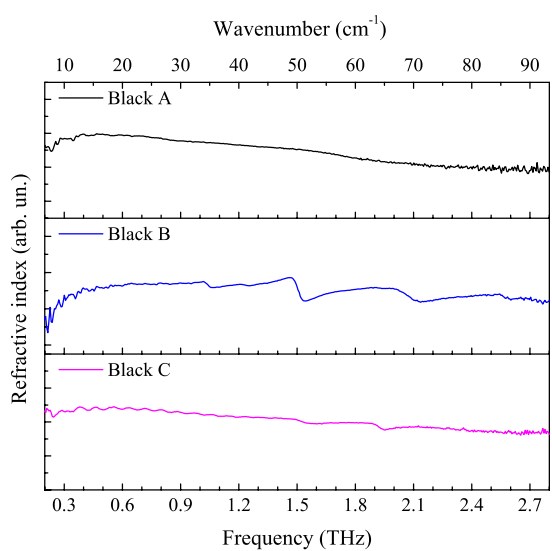

Figure 5. Frequency behaviour of the refractive indices for the iron gall inks measured on pellets. The dispersive features of the refractive index for the Black B and Black C inks confirm what measured in the respective absorbance spectra.

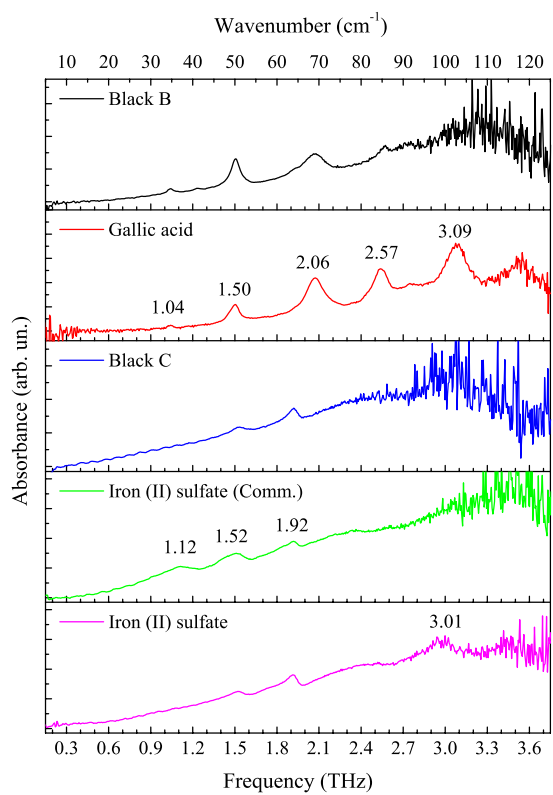

Figure 6. THz absorbance spectra for the Black B and Black C pigments together with those of the individual constituents gallic acid, iron(II) sulfate commercial, and iron(II) sulfate synthesized in lab.

## 4. CONCLUSIONS

THz time-domain spectroscopy in transmission configuration was applied to investigate inks commonly used in artworks, i.e. red, indigotin-based, iron-gall and lead white inks. Pellet samples were prepared, blending the bulk dried pigments with PE powder. Thin film samples are made by liquid inks layered on PE pellicles and subsequently dried. The THz-TDS technique achieved high signal-to-noise ratio for both pellet and layered inks, enabling a particularly meaningful analysis of the data. The THz spectra measured show similar features in both sample types, pellet and film.

In particular, the red cochineal carmine does not show any evident spectral signature in the 0.1-3.2 THz range, whereas the red ochre shows a weak but distinct absorption peak at about 2.2 THz. The blue ink presents two weak absorption peaks at 1.5 and 1.95 THz. The absorbance spectrum of white lead shows a shallow feature at around 2 THz.

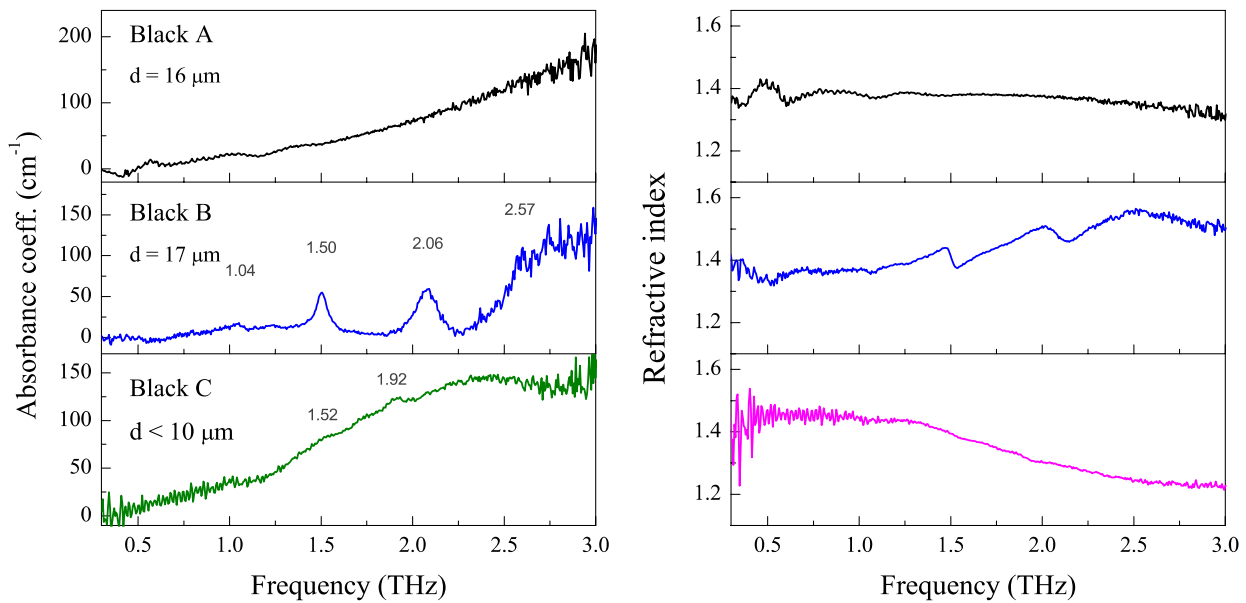

Figure 7. Absorption coefficient (left panel) and refractive index (right panel) as a function of frequency for the black inks on PE pellicles.

The spectroscopic features of the iron gall inks are more articulated and pronounced; we want to stress how the inks prepared by the ancient recipe (recipe A), can be immediately distinguished from those prepared with synthetic materials (recipe B) or commercial ones. In fact, the black inks prepared by recipe A show a featureless response. On the contrary, black inks prepared by recipe B or commercially available inks reveal the spectral signatures of their elementary constituents.

Moreover, the possibility to measure in absolute scale the THz spectra of layered inks brings these spectroscopic investigations closer to real-practice problems. In these respects, the present research is a rare attempt to improve the THz-TDS method in order to overcome the gap between the purely scientific uses and the operative applications in the field of cultural heritage.

## ACKNOWLEDGEMENT

This work was funded by Regione Toscana, prog. POR-CROFSE-UNIFI-26 and by Ente Cassa di Risparmio Firenze, prog. 2015-0857. The authors of INO-CNR acknowledge IPERION CH project GA 654028, funded by the EU community's H2020-research infrastructure program.